# 综合能源系统中最优潮流计算方法综述

皇甫夏亮

（华北水利水电大学电气工程学院，河南省 郑州市 450046）

## A review of the calculation methods of optimal power flow in integrated energy systems

HUANGPU Xia-liang

(School of Electrical Engineering， North China University of Water Resources and Electric Power，Zhengzhou City， Henan Province 450046， China)

**ABSTRACT:** The analysis of Integrated Energy Systems (IES) is crucial for enhancing the comprehensive and complementary utilization of clean energy across China, significantly impacting the effective planning, operational coordination, and security control of the IES network. This paper presents a systematic review of the current research on optimal power flow (OPF) within IES, addressing the spatiotemporal interrelationships and coupled co-supply among primary energy processes such as electricity, gas, and heat (cooling). It highlights the challenges and future directions in this field, underscoring the lack of comprehensive studies on coupled power flow modeling for electricity-heat-gas systems and the need for more robust modeling approaches that align with practical engineering applications. Furthermore, the paper discusses the potential of multi-target energy storage systems to enhance energy consumption efficiency and flexibility in energy resource management. The existing models and algorithms for target power flow and optimal power flow are critiqued for their lack of flexibility and comprehensiveness, particularly in handling multi-objective flows and ensuring system safety and reliability. The study emphasizes the necessity for further development of safety and reliability assessment frameworks to support the evolving demands of integrated energy systems.

**摘要：** 综合能源系统（IES）分析对促进清洁能源的综合互补利用及提高全国清洁能源利用效率具有关键意义，直接关系到 IES 网络的有效规划、运行协调及安全控制。本文系统回顾了 IES 中最优潮流（OPF）的研究现状，特别强调了电力、天然气与热能（冷能）等主要能源过程间的时空关系及联合供应的复杂性。文章指出，目前对于电-热-气耦合 IES 系统的潮流建模研究尚不充分，需要与实际工程应用更紧密地结合，发展更为健壮的模型方法。此外，探讨了多目标储能系统在提升能源消耗效率及资源调配灵活性方面的潜力。现有的目标潮流及最优潮流模型算法在处理多目标潮流和确保系统安全与可靠性方面存在限制，需进一步发展安全性和可靠性评估框架以支撑综合能源系统的发展需求。

## 0 引言

随着社会工业科技生产规模的持续发展规模与中国人民生活水平要求的层次不断地提高，能源消费与世界气候形势会日益地严峻，能源互联网具有开放、可持续等特点，是当前能源领域的一个重要发展方向[1-3]。为及时响应落实国家发展"双碳"战略政策任务的政策号召，研究建设集电、气、热等多种先进能源形式系统为于一体的综合能源系统(Integrated Energy System， IES)方案[4]。IES 可以同时有效实现提高社会能源利用率目标和改善供蓄能转换系统能源的运行整体安全性，并能实现目标"碳减排"[5-6]。

综合能源系统已经成为世界能源研究的一个重要的战略方向，也是世界上最重要的科技高地。欧洲率先提出了综合能源系统这一理念，并在欧盟框架计划下进行了多能协调优化及相关能源系统研究，欧洲国家也对此作出了积极响应[7-8]。美国在 2007 年通过了《能源自主与安全法案》，提出了实施全面能源规划的必要性[9]。加拿大致力于在 2050 年前实现碳排放减少，并将主要精力集中在社区集成能源系统（ICES）上，并计划在全国范围内进行大范围的推广[10-11]。日本作为亚洲第一个对能源系统进行研究的国家，在 2010 年发

布了《日本战略能源计划》，提出了以安全、环保、高效为目标的电力和天然气系统[12]。

目前我国 IES 领域的创新发展实践仍可能面临围绕着实现多尺度能源有效耦合、多时间尺度能源等两个方面提出的重要技术问题挑战[13]。目前，针对分布式 IES 应用的技术研究内容主要还是围绕分布式多功能流的耦合和建模、系统状态风险估计、安全性能分析建模与优化控制建模和网络优化控制调度建模展开。

潮流分析技术的核心概念最初源自现代电力系统，主要分析用于计算求解在电力网络环境中网络各传输节点对系统的输入电压分配和传输功率的分布，判断对其输入负荷调节的动态合理性程度并用于计算确定其传输损耗的大小[14]。如今，潮流耦合分析理论已经成功广泛被应用于 IES 多负能电流耦合分析计算平台，其流稳定理论建模工具与数值求解分析方法都已有奠定了较广泛深入的基础研究基础。文献[15-16]全面系统详细地系统总结分析了 IES 中各耦合子系统及各种耦合子元件变动的控制系统稳态模型描述与稳态计算求解方法，但结果表明仅是依靠控制系统稳态潮流变化规律不能较精确的刻画计算出耦合系统状况连续稳定变动中的系统动态过程模型[17]，且也不能准确地对应多种不确定性因素叠加(如电力负荷波动、设备故障或市场需求量变动等)所对耦合系统状况稳定的冲击效应模型[18]。文献［19］主要为分析能源系统的动态特性。文献［20］则提出了一种 IES 状态估计。鉴于综合能源系统的研究起步比较晚，近几年来，综合能源系统的功率流计算研究也基本以牛顿-拉夫逊法[21-22] 及其推广和完善为基础。当前潮流的相关分析还多停留根据稳态潮流，对各种异质低能流之间的动态时空尺度差异及动态不确定性差异方面的影响顾及到较少等实际问题，缺乏专门针对各种不同流工况潮流计算分析方法体系的体系归纳。而将最优潮流问题分析作为稳态潮流计算分析技术的进一步应用优化，在国际 IES 规范中得到的实际应用与发展均尚未充分成熟，仍主要需综合参考各种电力系统问题的求解方法。因此，需要更体系详细地全面总结对 IES 在不同工况阶段下的基本潮流预测和对最优潮流模型的分析。在此基础上，研究综合能源系统的能量流动规律，构建综合能源系统的安全性评价体系[23-25]，对于提升综合能源系统的能量供给与可靠性具有十分重要的意义。

为更好方便和研究在 IES 系统不同种类能源流间流的潮流相互转换机制与能量耦合及其关系，实现高效清洁能源有效利用分析与能源系统效率优化，本文重点总结介绍了 IES 最优潮流现状，同时，介绍 IES 最优潮流的问题和方法。最后，总结归纳目前的多能流预测分析及其研究各种模型手段与统计方法应用的种种局限性，并试图对国内外未来最优潮流与分析领域的若干研究热点方向提出若干建议探讨与研究展望。

## 1 IES

### 1.1 IES 定义

IES 是一个指在一定空间区域网络内利用国内外先进实用的电能信息传输管理调度技术，集电、热、天然气资源等多种主要能源物资的集中生产、输送管理与物资分配、转换、存储管理和运输消费服务等诸多环节服务于为一体，实现我国多种类型能源结构的综合有机统一协调管理与服务优化，促进社会可持续稳定发展能力的现代能源产供销一体化服务系统[6-7]。

### 1.2 变量和单位的要求

IES 通常包括多种能量供给网络系统（供电网、供冷网络、供气网络）、耦合链路、储能系统设备子系统以及负载等子系统[26]。IES 内部结构示意图如图 1 中所示。

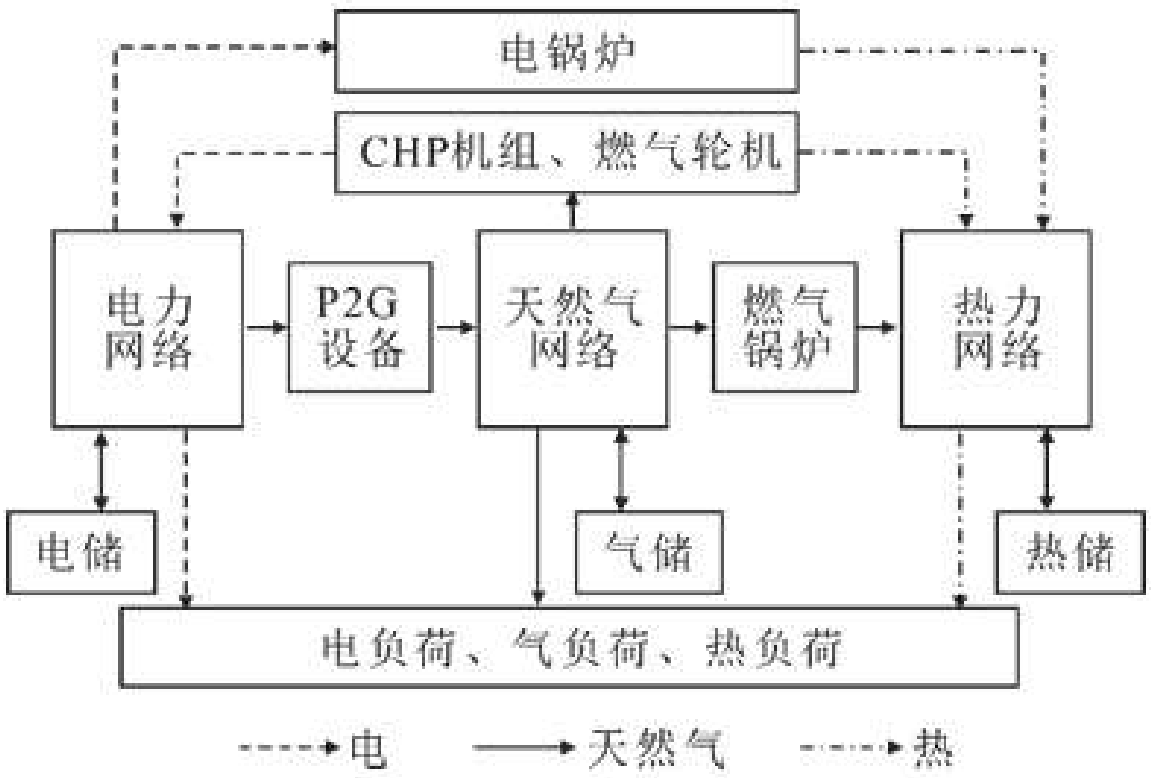

**图 1 IES 结构示意**

**Fig.1 The structure diagram of integrated energy system**

图 1 中：电力网络一般均含分布式光伏、风力发电机组或任何其他集中式可清洁再生化石能源的发电供应系统中;而热力网络是由集中供水

热源网络和供回水源网络而组成的系统[27-28]；天然气网络则由气源、供回气管道网络和气源压缩机等组成，压缩机以满足保证气源一定等级的最大供气压给力，维持天然气供给气压稳定;耦合能设备能够充分实现两种不同结构形式能源利用之间高效的相互转化效率;储能设备可为电储能、热解储能器和储气罐，能够长期维持更多的能源供应的系统动态的平衡，有利于显著提高整个系统动态运行方式的经济灵活性度和经济性。

## 2 IES 多能流计算模型

### 2.1 电网模型

IES 中，电力网络有交流、直流两种模型[29-32]。IES 中，电网交流稳态潮流模型功率平衡方程定义为

$$\begin{aligned}\Delta P_i &= P_i - U_i \sum_{j=1}^{n} U_j (G_{ij}\cos\theta_{ij} + B_{ij}\sin\theta_{ij}) \\ \Delta Q_i &= Q_i - U_i \sum_{j=1}^{n} U_j (G_{ij}\sin\theta_{ij} - B_{ij}\cos\theta_{ij})\end{aligned} \quad (2)$$

直流潮流模型可理解为一种对交流模型的一个高度简化，可以描述其为

$$\begin{aligned}&P_i = \sum_{j\in i} B_{ij}\theta_j, i = 1,2,\cdots,n \\ &B_{ij} = -\frac{1}{x_{ij}}, B_{ii} = \sum_{j\in i}\frac{1}{x_{ij}}, i \neq j\end{aligned} \quad (3)$$

### 2.2 热网模型

热网潮流模型[33-36]如式（4）所示

$$\begin{aligned}\boldsymbol{A}q_m &= q_{m,\mathrm{in}}, \\ \boldsymbol{M}h_{\mathrm{f}} &= 0,\end{aligned} \quad (4)$$

管道水头损失的计算方法为

$$h_{\mathrm{f}} = Kq_m|q_m| \quad (5)$$

热力方程中需考虑问题如式（6）所示

$$\begin{aligned}&\boldsymbol{\Phi} = c_{p,\mathrm{water}} q_m (T_{\mathrm{s}} - T_{\mathrm{o}}) \\ &T_{\mathrm{end}} = T_{\mathrm{start}} - T_\alpha \mathrm{e}^{-\frac{\lambda L}{c_{p,\mathrm{water}} q_m}} + T_\alpha \\ &\left(\Sigma q_{m,\mathrm{out}}\right) T_{\mathrm{out}} = \Sigma\left(q_{m,\mathrm{in}} T_{\mathrm{in}}\right)\end{aligned} \quad (6)$$

### 2.3 天然气网络模型

天然气网模型可再分 2 类模型。管道气体可以定常向前流动，动能系数的微小变化甚至可直接忽略和不计，无需单独使用压缩机[37]。则设其管道流量方程式为

$$\begin{aligned}&q_{m,mn} = K_{mn} s_{mn} \sqrt{s_{mn}(p_m^2 - p_n^2)} \\ &s_{mn} = \begin{cases} +1 & p_m - p_n \geqslant 0 \\ -1 & p_m - p_n < 0 \end{cases}\end{aligned} \quad (7)$$

然而在大多数使用情况条件下，天然气管道本身的机械摩擦与阻力并不可轻易忽视[38]，考虑压缩机的天然气网节点的流量平衡方程为

$$\sum_{n\in C_m} q_{m,mn} + \sum_{k\in C_m} q_{m,\tau k} = q_{m,\mathrm{gas}} \quad (8)$$

压缩机消耗的等效电能 $W_k$ 和流量 $q_{m,\tau k}$ 可表示为

$$\begin{aligned}&W_k = B_k q_{m,kmn}\left[\left(\frac{p_m}{p_n}\right)^{Z_k\left(\frac{r-1}{r}\right)} - 1\right] \\ &q_{m,\tau k} = K_{mn}\sqrt{s_{mn}(p_m^2 - p_n^2)}\end{aligned} \quad (9)$$

### 2.4 耦合设备模型

#### 2.4.1 CHP 机组模型

CHP 机组[39-40] 利用天然气燃烧产生的高品位余热产生余热，中低品位余热供暖，实现 IES 异质能量的耦合。

$$\begin{cases} c_{\mathrm{m}} = \dfrac{\phi_{\mathrm{CHP}}}{P_{\mathrm{CHP}}} \\ P_{\mathrm{CHP}} = \eta_{\mathrm{e}} G_{\mathrm{CHP}} \end{cases} \quad (10)$$

#### 2.4.2 P2H 设备模型

P2H 装置包含电锅炉、热泵和储热装置等[41]。电锅炉是一种可以将电力直接转换成热能的设备，它通常是供热系统的辅助热源，它参与了热、电负荷的峰谷调节。

$$\phi_{\mathrm{b}} = \eta_{\mathrm{b}} P_{\mathrm{b}} \quad (11)$$

热泵是一种与制冷压缩机相似的制冷系统，它是一种利用外界热能，并以较小的电能为热源的制冷系统。

$$\eta_{\mathrm{hp}} = \frac{\Phi_{\mathrm{hp}}}{P_{\mathrm{hp}}} \quad (12)$$

#### 2.4.2 P2G 设备模型

P2G 技术[42-45] 将电能转换成化学能，按照产品的不同，可以将其划分为两大类：一是电制氢，二是电制氢。电转氢是指通过电解水制氢的方式，在水中生成氢、氧，从而实现对氢的直接利用。然

而，传统的方法是将氢与$CO_2$进行反应，制备出以甲烷为原料的气体，这一过程被称为“电转气”，其产品可以直接进入输气管网，实现大规模储存与长距离输送。例如，将电力转换成天然气，它的数学模型可以表示为

$$G_{\mathrm{P2G}} = \frac{3600\eta_{\mathrm{P2G}}}{f_{\mathrm{LHV}}} P_{\mathrm{P2G}} \tag{13}$$

## 3 IES 最优潮流现状

### 3.1 问题描述

IES 优化潮流计算方法的扩展，实现多能互补与协调优化调度，对实现综合能源系统的经济、稳定、安全运行具有十分重要的意义。建立 IES 最优潮流模型时，必须假定：1）所有能量单元的工作状态都已知，不需要考虑单元间的组合；2) 根据实际的工作条件，确定了每个能量装置的输出；3) 综合能源系统的拓扑已经确定，没有发生线路变化、网络重配置等问题。

最优潮流的数学模型可表示为

$$\min f = f(x,u) \quad \text{s.t} \quad \begin{cases} g(x,u) = 0 \\ h(x,u) \leqslant 0 \end{cases} \tag{14}$$

IES 最优潮流问题是一个具有多重约束条件的复杂非线性优化问题，国际上对该问题的研究刚刚起步。鉴于综合能源系统的潮流优化与电力系统具有许多相似性，借鉴电力系统的方法，当前和将来可以拓展到综合能源系统的最优潮流计算方法有：1）经典算法，如线性规划，牛顿法等。2) 智能算法，例如：粒子群（PSO)，遗传算法（GA) 等。

### 3.2 经典算法

传统的综合电力系统最优潮流计算方法，如整数规划、内点法等，已经被初步运用到综合电力系统的最优潮流计算中，其方法简单，求解速度快，适合中、小型、单目标优化问题。然而，传统算法的编程难度大，收敛速度受初始值的影响，并且当目标函数为间断或者多个极值时，其收敛速度会受到限制，无法解决大规模、强非线性等问题。

文献[46]以综合能源系统的运行费用最小化为优化目标，以安全性与污染物排放为约束，利用内部点法求解综合能源系统的优化潮流，探索不同的目标与限制条件下，环境因子对综合能源系统的影响。文献[47]通过构建准确的热力系统稳态能量流量计算模型，以热电联产单元的电功率、电锅炉功率为调整变量，研究基于内点法的电能-热力综合能源系统能量流量优化算法。文献[48]拟构建具有多分枝辐射状供热系统的热-电耦合优化潮流数学模型，并基于内点算法的快速收敛和稳健性，对以上模型进行求解。

### 3.3 智能算法

智能算法通常是从人类智能、生物群体的社会性或者自然界的规律中提取出来的。本文介绍了一种基于神经网络的综合能源系统优化潮流计算方法。该类算法一般不要求目标函数或约束是连续或凸的，有的甚至无需解析形式，并且对计算结果的不确定性有较好的适应性。然而，目前智能优化算法的研究还不够完善，且往往无法保证求解结果的最优性，更多地将其看作是一种启发式算法。

文献[49]在能量中心原理的基础上，建立了一个热电联供系统的数学模型，并给出了一个集成的潮流计算方法。在此基础上，综合考虑微能源系统的运行特点，构建适用于微能源系统的最优潮流计算模型，并基于粒子群算法对其进行优化。文献[50]提出了一种新颖的双阶段多目标最优潮流算法，用于协调 VSC-HVDC 混合交直流电网的经济性、电压稳定性和环境需求。文献[51]为求解电动汽车综合能源系统的多目标优化问题，提出了一种新的基于目标值互换的分布式多目标模糊优化方法。文献[52] 将遗传算法应用于带能量枢纽的大型水电综合能源系统的优化潮流计算。文献[53]以能量枢纽为基础，构建综合能源网络最优潮流模型，并在此基础上，结合多智能体（MAGA）的多智能体演化算法（MAGA)，对其进行求解。

### 3.4 其他算法

部分研究综合了多能流对综合能源系统多能流的影响，并将其作为控制因子进行优化，以减少系统运行费用。或者，研究综合能源系统中综合能源系统的最优潮流，并给出可并行优化过程的求解方法。文献[54] 综合考虑了区域综合能源系统中的三相不对称配电网、燃气管网和能源中

心等因素，以最小化运营费用为优化目标，在多能流优化模型中引入配网重构能力。文献[55]提出了一个数据驱动的两阶段分布鲁棒优化模型，用于调度园区综合能源系统，有效协调需求响应和可再生能源发电的不确定性，提高能源系统的经济效率和鲁棒性。基于高斯赛德尔型交变乘子法（ADMM），构建交直流互联电网的分布式优化潮流计算模型，并基于此设计一种新型的、快速收敛和并行优化的新型并行算法。文献[56]以所提出的同步类型 ADMM 为基础，开展电-气综合能源系统的分布优化能流求解研究。此外，由于深度强化学习结合了深度学习的强大特征提取能力和强化学习的决策制定机制，能够在复杂的、高维度的环境中学习最优策略[57-58]，文献[59]提出了一种基于深度强化学习的交流最优潮流方法，考虑了可再生能源和拓扑变化的不确定性，有效地协助电网运营商做出有效的实时决策。

## 4 最优潮流计算的挑战与未来发展

目前对 IES 的潮流研究及最优潮流方法研究应用尚属处于研究基础及发展研究阶段，国内外多数学者也已努力抓住了多可能流分析技术的几个基本理论方向，但同时还仍存在着以下诸多问题与不足:

1) 关于电-热或电热-热气 IES 模型的多能流建模与算法的研究报道较多[60]， 缺乏对于电热-热电-电热气 IES 建模的全面深入系统探究。

2) 国内目前进行的能源 IES 潮流的研究也多以采用稳态模型进行开展，未能够充分地全面考察各种异质能源系统模型相互之间实际存在较大的时间尺寸差别，准稳态模型及暂态模型相关研究数量较少[61]。

3) 国内现有的多能流计算模型也大多仅为一种弱电耦合计算模型，未真正全面系统地考察了各种能量间的电磁相互作用; 能源集线器理论尽管还可以统一和描述各耦合的环节，但对于其所未计及的耦合元件的各种非线性耦合和动态耦合等物理特性，应尽量根据工程实际工作情况去深化各个耦合的环节耦合的数值建模方法[62-64]。

4) 根据 IES 模型配置的多元储能装置能够帮助进一步设计提高电网能效，有利于分布式能源管理系统的快速灵活和调用[65-66]，但目前的研究模型都很少计及多元储能的装置。

5) 随着电力能源系统数字化转型的深入发展，系统面临的信息安全威胁与日俱增，特别是信息攻击对系统的潜在破坏效果不容忽视[67-69]。例如，虚假数据注入攻击不仅能够篡改关键运行数据或交易信息，还可能导致系统运行异常，严重威胁到电力系统的安全经济运行[70-72]。开展针对信息攻击具有韧性的 IES 最优潮流研究对增强系统安全和保障能源供应连续性具有重要意义，亟待加强。

6) 基于 IES 标准建立的安全及可靠性能评估框架体系仍有待企业进一步建立完善。此外，随着可再生能源及其不确定性的增加，系统的脆弱性进一步加剧[73-75]。计再生能源及可利用再生清洁能源出力、需求侧响应时间等不确定性因素引发的大概率潮流的研究更需逐步深化[76-77]。

经过充分调研及分析，IES 潮流计算模型可以用在对以下四个方面问题作了进一步研究。

1) 我国当前燃气网的潮流模型目前主要都以石油天然气为重点研究的对象，而目前对许多其他能源形式下的混合燃气网络进行研究得较少，面对此种气网潮流需要重新分析校核出其各主要节点之间的气压力[78]。

2) 由于目前动态潮流模型体系间的相互作用深度普遍不十分高，大多只简单考虑到了这 2 个联合子系统; 且并未特别考虑到各耦合器设备自身的动态特性; 因此目前对涉及到多耦合系统、多耦合设备问题的 IES 模型问题仍远需在后续的研究进一步深化[79-80]。此外，由于目前仍普遍所采用传统的差分法而难以准确平衡出计算余量大小与保证计算结果精度，需要逐步研究设计出一些更准确有效的计算模型方法与一种更高效节能的计算方法。

3) 在不确定性趋势模型中，概率潮流模型要求提供巨量的样品量来支撑来充分保证了其模型正确性[81]， 而时间趋势分析方法要求提供的样品量相对虽更少，却可能也就存在大概率数据遗漏的现象[82]。因此也可以进一步考虑怎样将以上这 2 种技术方法两者的比较优势特性相结合，应用于各种实际应用工程中。

4) 在模拟电力系统问题的研究 OPF 中， 模拟退火法仍是较重要的人工智能问题求解研究方法类型之一，而对 IES 中关于此种求解方法类型

的相关研究尚存在理论空缺[83-84]。

综上所述，未来的 IES 规模可能更大、耦合可能更强，统一能路理论、多功能流的算法的扩展可能仍是后续研究方向的一大重难点。此外，IES 技术中的技术潮流与研究创新还尚需通过耦合环节的建模、安全风险评估、不确定性的分析评价等创新工作手段来持续推动，以确保加快推动建设安全清洁运行可靠、调控科学智能、管理科学先进适用的 IES。

## 5 结论

本文综述了综合能源系统中最优潮流的计算方法，全面分析了当前的研究现状、面临的挑战以及未来的发展方向。文章详细探讨了电力、热能和天然气等多能源系统的时空耦合与联合供应问题，强调了在现有 IES 模型中对这些能源间动态互动和耦合效应的建模不足。研究指出，现有的潮流建模方法尚未能有效地反映出可再生能源高渗透和能源负荷变化带来的系统复杂性。因此，文章呼吁开发更为鲁棒的建模技术，以更好地处理多能源系统的非线性和时变特性，并应对可再生能源的随机性。本综述还强调了多目标储能系统在提升系统操作灵活性和能效方面的潜力，推动了在 OPF 框架中更广泛地应用决策支持工具，以增强系统对信息安全威胁和物理故障的韧性。通过这些改进，能源系统的管理和运行将更加高效和可靠，从而推动能源系统朝着更加可持续和环保的方向发展，旨在为实现更高效、安全和绿色的能源利用提供科学依据和技术支持。

## 参考文献